\documentstyle[preprint,tighten,epsf,aps]{revtex}
\begin{document}

\title{Observation of the Top Quark}

%
\author{
S.~Abachi,$^{12}$
B.~Abbott,$^{33}$
M.~Abolins,$^{23}$
B.S.~Acharya,$^{40}$
I.~Adam,$^{10}$
D.L.~Adams,$^{34}$
M.~Adams,$^{15}$
S.~Ahn,$^{12}$
H.~Aihara,$^{20}$
J.~Alitti,$^{36}$
% ...
\\
\vskip 0.50cm
\centerline{(D\O\ Collaboration)}
\vskip 0.50cm
}
\address{
\centerline{$^{1}$Universidad de los Andes, Bogota, Colombia}
\centerline{$^{2}$University of Arizona, Tucson, Arizona 85721}
\centerline{$^{3}$Brookhaven National Laboratory, Upton, New York 11973}
\centerline{$^{4}$Brown University, Providence, Rhode Island 02912}
\centerline{$^{5}$University of California, Davis, California 95616}
\centerline{$^{6}$University of California, Irvine, California 92717}
\centerline{$^{7}$University of California, Riverside, California 92521}
\centerline{$^{8}$LAFEX, Centro Brasileiro de Pesquisas F{\'\i}sicas,
                  Rio de Janeiro, Brazil}
\centerline{$^{9}$CINVESTAV, Mexico City, Mexico}
\centerline{$^{10}$Columbia University, New York, New York 10027}
\centerline{$^{11}$Delhi University, Delhi, India 110007}
\centerline{$^{12}$Fermi National Accelerator Laboratory, Batavia,
                   Illinois 60510}
\centerline{$^{13}$Florida State University, Tallahassee, Florida 32306}
\centerline{$^{14}$University of Hawaii, Honolulu, Hawaii 96822}
\centerline{$^{15}$University of Illinois, Chicago, Illinois 60680}
\centerline{$^{16}$Indiana University, Bloomington, Indiana 47405}
\centerline{$^{17}$Iowa State University, Ames, Iowa 50011}
\centerline{$^{18}$Korea University, Seoul, Korea}
\centerline{$^{19}$Kyungsung University, Pusan, Korea}
\centerline{$^{20}$Lawrence Berkeley Laboratory, Berkeley, California 94720}
\centerline{$^{21}$University of Maryland, College Park, Maryland 20742}
\centerline{$^{22}$University of Michigan, Ann Arbor, Michigan 48109}
\centerline{$^{23}$Michigan State University, East Lansing, Michigan 48824}
\centerline{$^{24}$Moscow State University, Moscow, Russia}
\centerline{$^{25}$University of Nebraska, Lincoln, Nebraska 68588}
\centerline{$^{26}$New York University, New York, New York 10003}
\centerline{$^{27}$Northeastern University, Boston, Massachusetts 02115}
\centerline{$^{28}$Northern Illinois University, DeKalb, Illinois 60115}
\centerline{$^{29}$Northwestern University, Evanston, Illinois 60208}
\centerline{$^{30}$University of Notre Dame, Notre Dame, Indiana 46556}
\centerline{$^{31}$University of Panjab, Chandigarh 16-00-14, India}
\centerline{$^{32}$Institute for High Energy Physics, 142-284 Protvino, Russia}
\centerline{$^{33}$Purdue University, West Lafayette, Indiana 47907}
\centerline{$^{34}$Rice University, Houston, Texas 77251}
\centerline{$^{35}$University of Rochester, Rochester, New York 14627}
\centerline{$^{36}$CEA, DAPNIA/Service de Physique des Particules, CE-SACLAY,
                   France}
\centerline{$^{37}$Seoul National University, Seoul, Korea}
\centerline{$^{38}$State University of New York, Stony Brook, New York 11794}
\centerline{$^{39}$SSC Laboratory, Dallas, Texas 75237}
\centerline{$^{40}$Tata Institute of Fundamental Research,
                   Colaba, Bombay 400005, India}
\centerline{$^{41}$University of Texas, Arlington, Texas 76019}
\centerline{$^{42}$Texas A\&M University, College Station, Texas 77843}
}

\date{\today}

\maketitle

\vspace{-0.3in}

\begin{abstract}
The D\O\ collaboration reports on a search for the Standard Model top
quark in $p\bar p$ collisions at $\sqrt{s} = 1.8$~TeV at the Fermilab
Tevatron, with
an integrated luminosity of approximately 50~pb$^{-1}$.
We have searched
for $t\bar t$ production in the dilepton
and single-lepton decay channels, with and without tagging of
$b$-quark jets.
We observed 17 events with an expected background of
\hbox{$3.8\pm0.6$} events.  The probability for an upward fluctuation of the
background to produce the observed signal
is $2\times10^{-6}$ (equivalent to 4.6 standard deviations).
The kinematic properties of the excess events are consistent with
top quark decay.
We conclude that we have observed the top quark and measure its mass
to be $199^{+19}_{-21}$ (stat.)~$\pm 22$ (syst.)~GeV/c$^2$ and its
production cross section to be $6.4 \pm 2.2$~pb.
\end{abstract}





In the Standard Model (SM), the top quark is the weak isospin
partner of the $b$ quark.  The D\O\ collaboration published a lower
limit on the mass of the top
quark of $131 \mbox{ GeV/c}^2$, at a confidence level (CL) of
95\%, based on an integrated luminosity of
13.5~pb$^{-1}$~\cite{boazprl}.
A subsequent publication~\cite{herbprl} reported the top quark
production cross section as a
function of the assumed top quark mass.
In that analysis, we found nine events with
an expected background of $3.8 \pm 0.9$ events
(statistical significance 1.9 standard deviations) corresponding to a
production cross section of $8.2\pm5.1$~pb for an assumed top quark mass of
180~GeV/c$^2$.
The CDF collaboration published evidence for top quark production
with a statistical significance of 2.8 standard deviations, a top quark
of mass $174 \pm 10^{+13}_{-12} \mbox{ GeV/c}^2$, and a production
cross section of $13.9^{+6.1}_{-4.8}$~pb~\cite{cdfnewtop}.
Precision electroweak measurements predict
a SM top quark mass of approximately 150--210~GeV/c$^2$, depending on the
mass of the Higgs boson~\cite{lepew}.
In the present paper, we report new results from the D\O\ experiment that
firmly establish the existence of the top quark.

We assume that the top quark is pair-produced and decays according to the
minimal SM ({\it i.e.}\ $t\bar t \rightarrow W^+ W^-b\bar b$).
We have searched for the top quark in channels where both $W$ bosons
decayed leptonically ($e \mu + \rm{jets}$, $e e + \rm{jets}$, and
$\mu \mu + \rm{jets}$) and in channels where just one $W$ boson decayed
leptonically ($e +\rm{jets}$ and $\mu +\rm{jets}$).
The single-lepton channels were subdivided into $b$-tagged and
untagged channels according to whether or not a muon was observed
consistent with $b\to\mu + X$.  The muon-tagged channels are
denoted $e +\rm{jets}/\mu$ and $\mu +\rm{jets}/\mu$.

Here we present an analysis
based on data collected at the Fermilab Tevatron at $\sqrt{s} = 1.8$~TeV
with an integrated luminosity of 44--56~pb$^{-1}$,
depending on the channel.
In the present analysis, the signal-to-background ratio for a high mass
top quark was substantially improved relative to Ref.~\cite{herbprl}.  An
optimization of the selection criteria was carried out using Monte Carlo
top quark events for signal and our standard background estimates.  The result
of this procedure was a factor of 3.7 better background rejection while
retaining 70\% of the acceptance for 180~GeV/c$^2$ top quarks.  This
corresponds to a signal-to-background ratio of 1:1 for a top quark mass
of 200~GeV/c$^2$, assuming the expected SM top cross
section~\cite{laenen}.  The improved rejection arises primarily by
requiring events to have a larger total transverse energy.



The D\O\ detector and data collection systems are described
in Ref.~\cite{dzeronim}.  The triggers and reconstruction algorithms for
jets, electrons,  muons, and neutrinos were the same as
those used in our previous top quark searches~\cite{boazprl,herbprl}.

The signature for the dilepton channels was defined as two
isolated leptons, at least two jets, and large missing transverse
energy $\rlap{\kern0.25em/}E_T$.
The signature for the single-lepton channels was defined as one
isolated lepton, large $\rlap{\kern0.25em/}E_T$, and a
minimum of three jets (with muon tag) or four jets (without tag).
The minimum transverse momentum $p_T$ of tagging muons was 4~GeV/c.
Requirements pertaining to the magnitude and direction of the
$\rlap{\kern0.25em/}E_T$, the aplanarity of the jets $\cal A$, and
the allowed ranges of pseudorapidity $\eta$ were similar to
Ref.~\cite{herbprl}.  Muons were restricted to $|\eta| < 1$ for
the last 70\% of the data because of forward muon chamber aging.
Events in the
$\mu\mu + {\rm jets}$ and $\mu +\rm{jets}/\mu$ channels were required to
be inconsistent with the $Z + \rm jets$ hypothesis, based on a global
kinematic fit.
The principal difference between the present analysis and the analysis of
Ref.~\cite{herbprl} was the imposition of a minimum requirement in all
channels on a quantity $H_T$, which we defined as the
scalar sum of the transverse energies $E_T$ of the jets
(for the single-lepton and $\mu\mu$ +
jets channels) or the scalar sum of the $E_T$'s of the leading electron
and the jets (for the $e\mu$~+~jets and $ee$~+~jets channels).
The kinematic requirements for our standard event selection
for all seven channels are summarized in Table~\ref{etht}.
In addition to the standard selection, we defined a set of loose event
selection requirements, which differed from the standard set by the
removal of the $H_T$ requirement and by the
relaxation of the aplanarity requirement for $e + \rm jets$ and
$\mu + \rm jets$ from ${\cal A} > 0.05$ to ${\cal A} > 0.03$.




For the dilepton channels, the main backgrounds were from
$Z$ and continuum Drell-Yan production
($Z,\gamma^* \rightarrow e e, \mu \mu$, and $\tau\tau$), vector boson pairs
($W W$, $W Z$), heavy flavor ($b\bar b$ and $c\bar c$) production,
and backgrounds with jets misidentified as leptons.
For the single-lepton channels, the main backgrounds were from
$W + \rm jets$, $Z + \rm jets$, and multijet production with a jet
misidentified as a lepton.
The method for estimating these backgrounds was the same as in our
previously published analyses~\cite{boazprl,herbprl}.

\begin{figure}
\epsfxsize=6.5in
\epsfbox{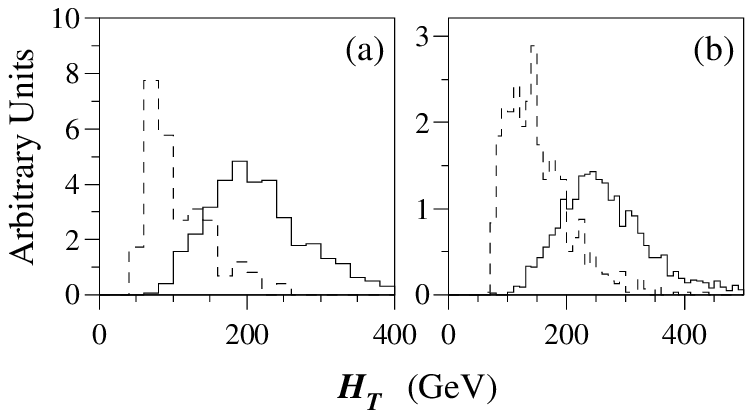}
\caption{Shape of $H_T$ distributions expected for the principal
backgrounds
(dashed line) and 200~GeV/c$^2$ top quarks (solid line) for
(a) $e\mu + \rm jets$ and
(b) untagged single-lepton + jets.}
\label{fig_ht1}
\end{figure}

\begin{figure}
\epsfxsize=6.5in
\epsfbox{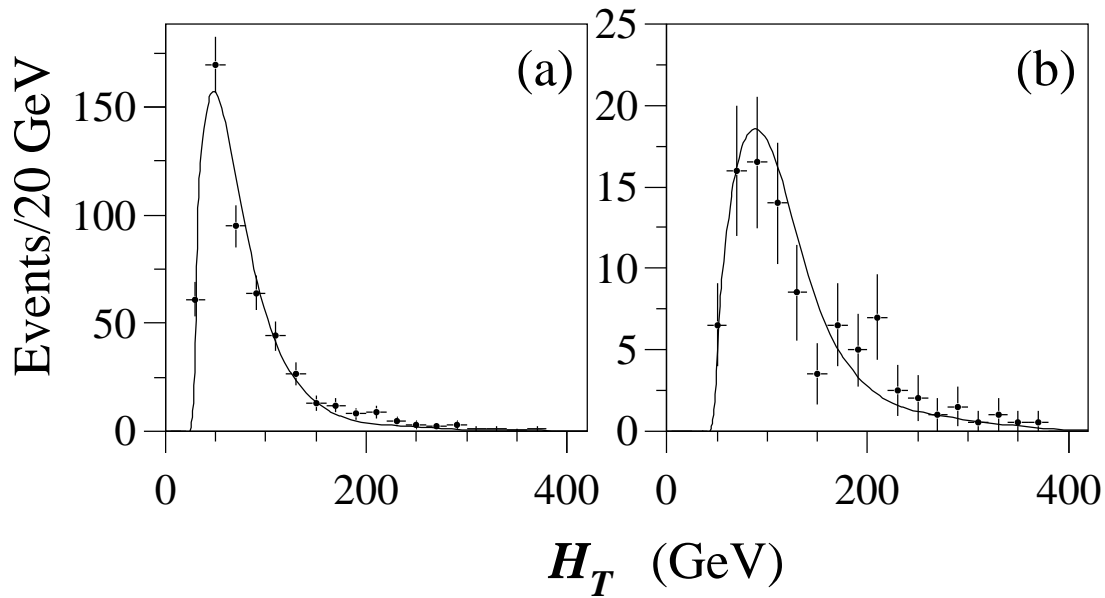}
\caption{Observed $H_T$ distributions (points) compared to
the distributions expected from background (line) for
$\rlap{\kern0.25em/}E_T > 25$~GeV/c and
(a) $e~+ \ge 2$ jets and (b) $e~+ \ge 3$ jets.}
\label{fig_ht2}
\end{figure}

$H_T$ is a powerful discriminator between background
and high-mass top quark production.  Figure~\ref{fig_ht1} shows a
comparison of the shapes of the $H_T$ distributions expected from
background and 200~GeV/c$^2$ top quarks in the channels
(a) $e\mu + {\rm jets}$ and (b) untagged single-lepton + jets.
We have tested our understanding of background $H_T$ distributions by
comparing data and calculated background in background-dominated
channels such as electron + two jets and electron + three jets (Fig.
{}~\ref{fig_ht2}).
The observed $H_T$ distribution agrees with the background
calculation, which includes contributions from both
$W + \rm jets$ as calculated by the {\sc vecbos} Monte
Carlo~\cite{vecbos} and multijet events.



The acceptance for $t\bar t$ events was calculated
using the {\sc isajet} event
generator~\cite{isajet} and a detector simulation based on the {\sc geant}
program~\cite{geant}.  As a check, the acceptance was also calculated
using the {\sc herwig} event generator~\cite{herwig}.  The difference
between {\sc isajet} and {\sc herwig}
was included in the systematic error.

\begin{figure}
\epsfxsize=6.5in
\epsfbox{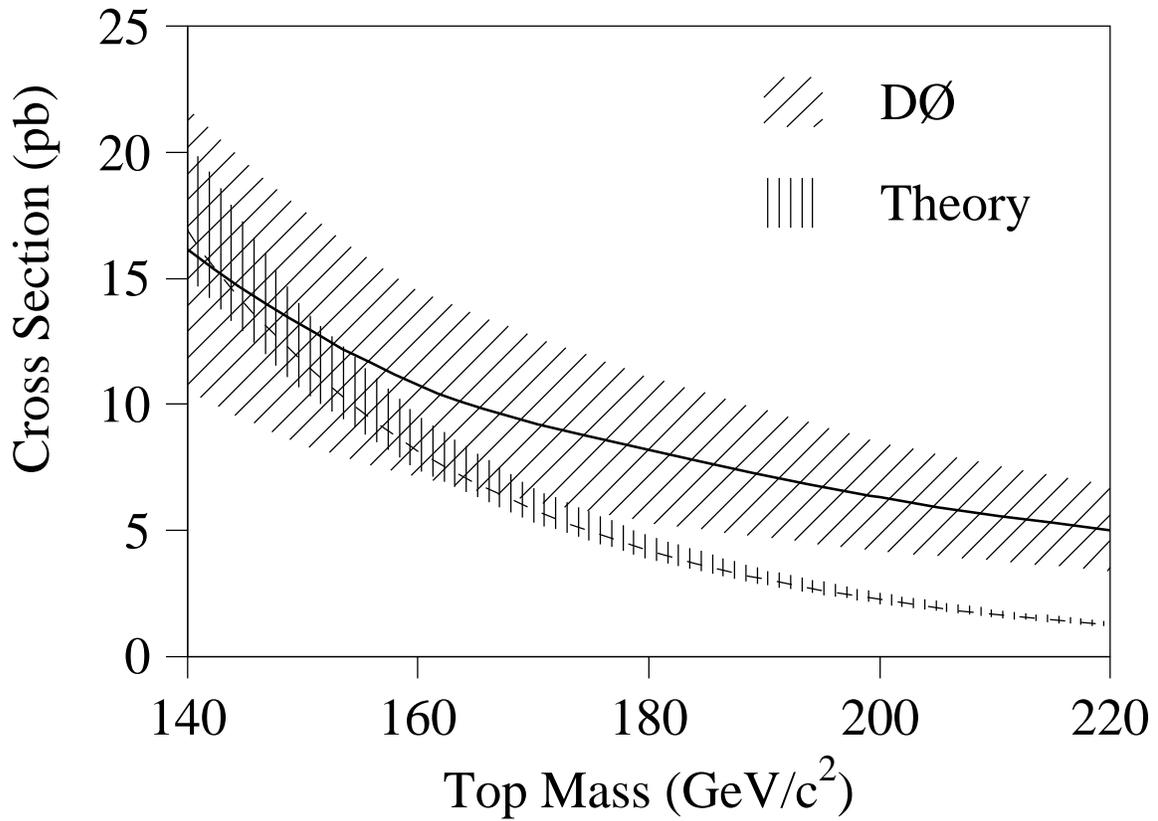}
\caption{D\O\ measured $t\bar t$ production cross section (solid line
with one standard deviation error band) as a function of assumed top
quark mass.  Also shown is the theoretical
cross section curve (dashed line)~\protect\cite{laenen}.}
\label{fig_xsec}
\end{figure}

{}From all seven channels, we observed 17 events with
an expected background of $3.8\pm0.6$ events (see Table~\ref{all}).
Our measured cross
section as a function of the top quark mass hypothesis is shown in
Fig.~\ref{fig_xsec}.  Assuming a top quark mass of 200~GeV/c$^2$,
the production cross section is $6.3\pm2.2$~pb.  The error in the cross
section includes an overall 12\% uncertainty in the luminosity.
The probability
of an upward fluctuation of the background to 17 or more events
is $2\times 10^{-6}$, which corresponds to 4.6 standard deviations
for a Gaussian probability distribution.  We have calculated
the probability for our observed distribution of excess events among the
seven channels and find that our results are consistent with top quark
branching fractions at the 53\% CL.  Thus, we observe a
statistically
significant excess of events and the distribution of events among
the seven channels is consistent with top quark production.





\begin{figure}
\epsfxsize=6.5in
\epsfbox{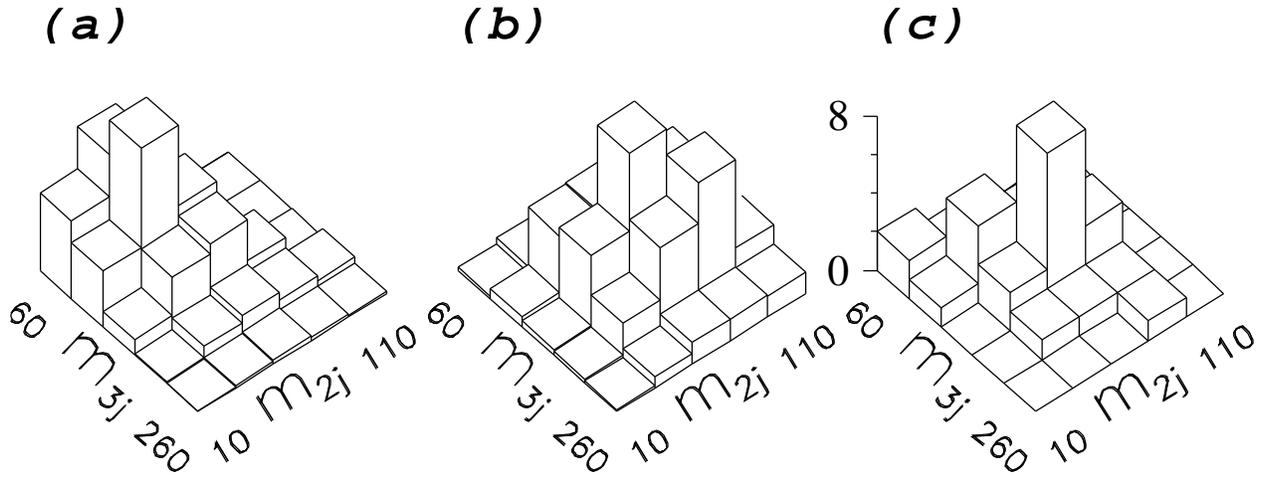}
\caption{Single-lepton + jets two-jet {\it vs.} three-jet invariant
mass distribution for
(a) background, (b) 200~GeV/c$^2$ top Monte Carlo ({\sc isajet}), and
(c) data.}
\label{fig_2d_mass}
\end{figure}

Additional confirmation that our observed excess contains a high-mass
object comes
from the invariant masses of jet combinations in single-lepton + jets
events.  For this analysis, we selected single-lepton + four-jet
events using the loose event selection requirements (27 events).  An
invariant mass analysis was performed, based on
the hypothesis $t\bar t\to W^+W^-b\bar b\to\ell\nu q\bar qb\bar b$.
One jet was assigned to the semileptonically decaying top quark
and three jets were assigned to the hadronically decaying
top quark.  The jet assignment algorithm attempted to assign one of the
two highest $E_T$ jets to the semileptonically decaying top quark and to
minimize the difference between the masses of the two top quarks.
The invariant mass of the three jets assigned to the hadronically
decaying
top quark is denoted by $m_{3\rm j}$.  The invariant mass of the pair
of hadronically decaying top quark jets with the smallest invariant mass
is denoted by
$m_{2\rm j}$.  Figure~\ref{fig_2d_mass} shows the
distribution of $m_{3\rm j}$ {\it vs.} $m_{2\rm j}$ for (a) background
($W + \rm jets$ and multijet)
(b) 200~GeV/c$^2$ top Monte Carlo, and (c) data.   The data are peaked at
higher invariant mass, in both dimensions, than the background.
Based only on the shapes of the distributions, the
hypothesis that the data are a combination of top quark
and background events (60\% CL) is
favored over the pure background hypothesis (3\% CL).





To measure the top quark mass, single-lepton + four-jet events
were subjected to 2-constraint kinematic fits to the hypothesis
$t\bar t\to W^+W^-b\bar b\to\ell\nu q\bar qb\bar b$.
Kinematic fits were performed on all permutations of the jet assignments
of the four highest $E_T$ jets,
with the provision that muon-tagged jets were always assigned to a $b$-quark
in the fit.
A maximum of three permutations with $\chi^2<7$ (two degrees of freedom)
were retained, and a single
$\chi^2$-probability-weighted average mass (``fitted mass'')
was calculated for each event.
Monte Carlo studies using the {\sc isajet} and {\sc herwig} event
generators showed that the fitted mass
was strongly correlated with the top quark mass.
Gluon radiation, jet assignment combinatorics, and the event selection
procedure introduced a shift in the fitted mass
(approximately $-20$~GeV/c$^2$ for 200~GeV/c$^2$ top quarks), which was taken
into account in the final mass determination.

\begin{figure}
\epsfxsize=6.5in
\epsfbox{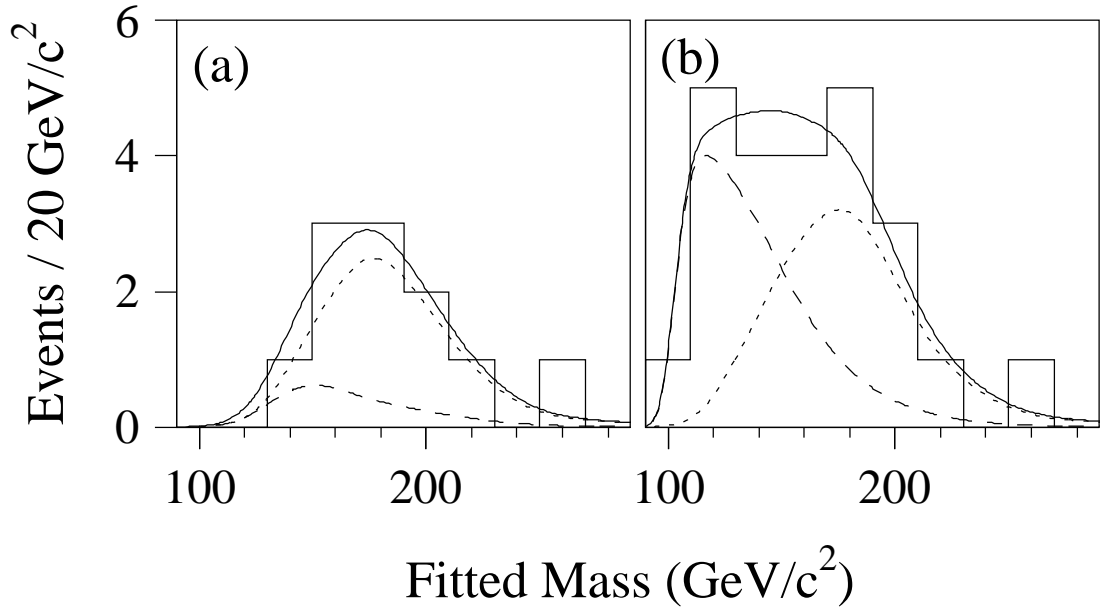}
\caption{Fitted mass distribution for candidate events (histogram) with the
expected mass distribution for 199~GeV/c$^2$
top quark events (dotted curve), background
(dashed curve), and the sum of top and background (solid curve) for
(a) standard and (b) loose event selection.}
\label{fig_mass}
\end{figure}

Eleven of the 14 single-lepton + jets candidate events selected
using the standard cuts were fitted successfully.
Figure~\ref{fig_mass}(a) shows the fitted mass distribution.
An unbinned likelihood fit,
incorporating top quark and background contributions, with the top quark
mass allowed to vary, was performed on the fitted mass distribution.
The top quark contribution was modeled using {\sc isajet}.
The background contributions were constrained to be consistent with
our background estimates.
The likelihood fit yielded a top quark mass of
$199^{+31}_{-25}$ (stat.)~GeV/c$^2$ and described the data well.

To increase the statistics available for the mass fit, and to remove
any bias from the standard $H_T$ requirement, we repeated the mass
analysis on events selected using the loose requirements.  Of 27
single-lepton + four-jet events, 24 were fitted successfully.
The removal of the $H_T$ requirement introduced a substantial background
contribution at lower mass in addition to the top signal, as shown in
Fig.~5(b).
A likelihood fit to the mass distribution resulted in a top quark
mass of $199^{+19}_{-21}$ (stat.)~GeV/c$^2$, consistent with the
result obtained from the standard event selection.  The result of the
likelihood fit did not depend significantly on whether the normalization
of the background was constrained.
Using {\sc herwig} to model the top quark contribution
resulted in a mass
4~GeV/c$^2$ below that found using {\sc isajet}.  This effect
was included in
the systematic error.  The total systematic error in the top quark mass is
22~GeV/c$^2$, which is dominated by the
uncertainty in the jet energy scale.



In conclusion, we report the observation of the top
quark.  We measure the top quark mass to be
$199^{+19}_{-21}$ (stat.)~$\pm 22$ (syst.)~GeV/c$^2$ and measure
a production cross section of $6.4 \pm 2.2$~pb at our central mass.


We thank the Fermilab Accelerator, Computing, and Research Divisions, and
the support staffs at the collaborating institutions for their contributions
to the success of this work.   We also acknowledge the support of the
U.S. Department of Energy,
the U.S. National Science Foundation,
the Commissariat \`a L'Energie Atomique in France,
the Ministry for Atomic Energy and the Ministry of Science and
Technology Policy in Russia,
CNPq in Brazil,
the Departments of Atomic Energy and Science and Education in India,
\pagebreak
Colciencias in Colombia, CONACyT in Mexico,
the Ministry of Education, Research Foundation and KOSEF in Korea,
and the A.~P.~Sloan Foundation.



\pagebreak

\begin{table}
\caption{Minimum kinematic requirements for the standard event selection
(energy in~GeV).}
\label{etht}
\begin{tabular}{c|c|c|c|c|c|c|c}
& \multicolumn{2}{c|}{Leptons} & \multicolumn{2}{c|}{Jets} & & & \\
\cline{2-5}
Channel & $E_T(e)$ & $p_T(\mu)$ & $N_{\rm jet}$ & $E_T$
& $\rlap{\kern0.25em/}E_T$ & $H_T$ & $\cal A$ \\
\hline
\hline
$e\mu$ + jets & 15 & 12 & $2$ & 15 & 20 & 120 & - \\
$ee$ + jets & 20 & & $2$ & 15 & 25 & 120 & - \\
$\mu\mu$ + jets & & 15 & $2$ & 15 & - & 100 & - \\
\hline
$e + {\rm jets}$ & 20 & & $4$ & 15 & 25 & 200 & 0.05 \\
$\mu + {\rm jets}$ & & 15 & $4$ & 15 & 20 & 200 & 0.05 \\
\hline
$e + {\rm jets}/\mu$ & 20 & & $3$ & 20 & 20 & 140 & - \\
$\mu + {\rm jets}/\mu$ & & 15 & $3$ & 20 & 20 & 140 & - \\
\end{tabular}
\end{table}


\begin{table*}
\squeezetable
\caption{Efficiency $\times$ branching fraction
($\varepsilon \times {\cal B}$) using standard event selection and
the expected number of top quark events ($\langle N \rangle$)
in the seven channels, based on the central theoretical $t\bar t$
production cross section of Ref.~\protect\cite{laenen}, for four top masses.
Also given are the
expected background, integrated luminosity, and the number of
observed events in each channel.}
\label{all}
\begin{tabular}[p]{cc|c|c|c|c|c|c|c|c}
\multicolumn{2}{c|}{$m_t$ (GeV/$c^2$)} &
\multicolumn{1}{c|}{$e \mu + \rm{jets}$} &
\multicolumn{1}{c|}{$e e + \rm{jets}$~~} &
\multicolumn{1}{c|}{$\mu \mu + \rm{jets}$~~~} &
\multicolumn{1}{c|}{$e + \rm jets$~} &
\multicolumn{1}{c|}{$\mu+\rm jets$~ } &
\multicolumn{1}{c|}{$e + \rm jets /\mu$} &
\multicolumn{1}{c|}{$\mu+\rm jets /\mu$} &
\multicolumn{1}{c}{ALL} \\
\hline
\hline
      & $\varepsilon \times {\cal B} (\%) $ & $0.17\pm 0.02$ & $0.11\pm 0.02$
      & $0.06\pm 0.01$ & $0.50\pm 0.10$ & $0.33\pm 0.08$ &  $0.36\pm 0.07$ &
        $0.20\pm 0.05$ & \\
            \cline{2-10}
 ~140 & $\langle N \rangle$             & $1.36\pm 0.21$ & $1.04\pm 0.19$
      & $0.46\pm 0.08$ & $4.05\pm 0.94$ & $2.47\pm 0.68$  &   $2.93\pm 0.68$
      & $1.48\pm 0.42$ & $13.80\pm 2.07$ \\
\hline
      & $\varepsilon \times {\cal B} (\%) $ & $0.24\pm 0.02$ & $0.15\pm 0.02$
      & $0.09\pm 0.02$ & $0.80\pm 0.10$ & $0.57\pm 0.13$ &  $0.50\pm 0.08$ &
        $0.25\pm 0.06$ & \\
            \cline{2-10}
 ~160 & $\langle N \rangle$             & $0.94\pm 0.13$ & $0.69\pm 0.12$
      & $0.34\pm 0.07$ & $3.13\pm 0.54$ & $2.04\pm 0.53$  &   $1.95\pm 0.39$
      & $0.92\pm 0.24$ & $10.01\pm 1.41$ \\
\hline
      & $\varepsilon \times {\cal B} (\%) $ & $0.28\pm 0.02$ & $0.17\pm 0.02$
      & $0.10\pm 0.02$ & $1.20\pm 0.30$ & $0.76\pm 0.17$ &  $0.56\pm 0.09$ &
        $0.35\pm 0.08$ & \\
            \cline{2-10}
 ~180 & $\langle N \rangle$             & $0.57\pm 0.07$ & $0.40\pm 0.07$
      & $0.19\pm 0.04$ & $2.42\pm 0.67$ & $1.41\pm 0.36$  &   $1.14\pm0.22$ &
        $0.64\pm 0.16$ & $6.77\pm 1.09$ \\
\hline
      & $\varepsilon \times {\cal B} (\%) $ & $0.31\pm 0.02$ & $0.20\pm 0.03$
      & $0.11\pm 0.02$ & $1.70\pm 0.20$ & $0.96\pm 0.21$ &  $0.74\pm 0.11$ &
        $0.41\pm 0.08$ & \\
            \cline{2-10}
 ~200 & $\langle N \rangle$             & $0.34\pm 0.04$ & $0.25\pm 0.05$
      & $0.11\pm 0.02$ & $1.84\pm 0.31$ & $0.95\pm 0.24$  &   $0.81\pm 0.16$ &
        $0.41\pm 0.10$ & $4.71\pm 0.66$ \\
\hline
\hline
\multicolumn{2}{c|}{Background}    & $0.12\pm 0.03$ & $0.28\pm 0.14$ &
$0.25\pm 0.04$ & $1.22\pm 0.42$ & $0.71\pm 0.28$ & $0.85\pm 0.14$  &
$0.36\pm 0.08$ & $3.79\pm 0.55$ \\
\hline
\multicolumn{2}{c|}{$\int {\cal L}dt \ (\rm pb^{-1})$} & $47.9\pm 5.7$ &
$55.7\pm 6.7$
& $44.2\pm 5.3$ & $47.9\pm 5.7$ & $44.2\pm 5.3$ & $47.9\pm 5.7$ &
$44.2\pm 5.3$ & \\
\hline
\multicolumn{2}{c|}{Data} & \multicolumn{1}{c|}{2} &
\multicolumn{1}{c|}{0} & \multicolumn{1}{c|}{1} &
\multicolumn{1}{c|}{5} & \multicolumn{1}{c|}{3} &
\multicolumn{1}{c|}{3} &  \multicolumn{1}{c|}{3} &
\multicolumn{1}{c}{17}
\end{tabular}
\end{table*}

\end{document}